# Complex solid solution electrocatalyst discovery by prediction and high-throughput experimentation

Thomas A. A. Batchelor[#,1], Tobias Löffler[#,2], Bin Xiao[#,3], Olga A. Krysiak[2], Valerie Strotkötter[3], Jack K. Pedersen[1], Christian M. Clausen[1], Alan Savan[3], Wolfgang Schuhmann*[,2], Jan Rossmeisl*[,1], Alfred Ludwig*[,3,4].

[1] Theoretical Catalysis – Center for High Entropy Alloy Catalysis (CHEAC), Department of Chemistry, University of Copenhagen, Copenhagen, Denmark.

[2] Analytical Chemistry – Center for Electrochemical Sciences (CES), Faculty of Chemistry and Biochemistry, Ruhr University Bochum, Universitätsst. 150, D-44780 Bochum, Germany.

[3] Chair for Materials Discovery and Interfaces, Institute for Materials, Faculty of Mechanical Engineering, Ruhr University Bochum, Universitätsst. 150, D-44780 Bochum, Germany.

[4] ZGH, Ruhr University Bochum, Universitätsst. 150, D-44780 Bochum, Germany.

[#] equal contributions

*Corresponding authors: Email: wolfgang.schuhmann@rub.de, jan.rossmeisl@chem.ku.dk , alfred.ludwig@rub.de

**Abstract**

Efficient discovery of electrocatalysts for electrochemical energy conversion reactions is of utmost importance to combat climate change. With the example of the oxygen reduction reaction we show that by utilising a data-driven discovery cycle, the multidimensionality challenge offered by compositionally complex solid solution ("high entropy alloy") electrocatalysts can be mastered. Iteratively refined computational models predict activity trends for quinary target compositions, around which continuous composition spread thin-film libraries are synthesized. High-throughput characterisation datasets are then input for refinement of the model. The refined model correctly predicts activity maxima of the exemplary model system Ag-Ir-Pd-Pt-Ru for the oxygen reduction reaction. The method can identify optimal complex solid solutions for electrochemical reactions in an unprecedented manner.

**Introduction**

Complex solid solutions (CSS) with five or more principal elements, often referred to as "high entropy alloys", were recently discovered to hold the promise of shifting the paradigm in materials design from 'using materials that we have' to 'engineering materials that we need'. In electrocatalysis there is an urgent need to discover new materials to catalyse the production of sustainable fuels and chemicals, needed for mitigating climate change. For many key reactions the scalability towards a global level is limited by present-day catalysts. However, most of the plausible billions of different materials have never been synthesized and nearly all materials that have been employed up to now represent edges and corners in the vast continuum of chemical space. Exploration of the promising multidimensional chemical space of CSSs with the aim of finding previously out-of-reach catalysts is extremely challenging, and requires an intelligent materials discovery strategy to focus on interesting regions of potential catalytic activity since neither simulation nor experiments alone can overcome serendipity in materials discovery.

The emerging paradigm change in electrocatalysis is based on the availability of millions of different active sites in the CSS atomic surface configuration. This unique arrangement of multiple elements directly neighbouring a binding site enables tuning activity by electronic and geometric effects. (*1, 2*) Recently, we discovered a noble-metal free CSS for the oxygen reduction reaction (ORR) (*3*), proposed a theoretical base for CSS catalysts as a discovery platform (*4*), applied this strategy on the $CO_2$ and CO reduction reactions (*5*), and described experimental challenges and concepts (*6*), which were confirmed experimentally (*7*). Moreover, the discovery and investigation of CSS electrocatalysts is attracting a burgeoning interest. Independently, several groups have shown that CSS electrocatalysts are indeed "game-changing" materials for a wide span of electrochemical reactions such as hydrogen and oxygen evolution reactions (*8–11*), CO (*5*), $CO_2$ (*5, 12*) and $O_2$ (*3, 4, 13*) reduction reactions as well as methanol oxidation (*14, 15*) or ammonia synthesis and decomposition (*16–18*). Outstanding activities were reported even though in most cases only the simplest case of equiatomic CSS compositions were investigated. This success was enabled by the development of an increasing number of mostly non-equilibrium synthesis methods comprising carbothermal shock synthesis (*16, 18*), combinatorial co-sputtering (*3, 7*), solvothermal reactions (*19*), ball milling (*12*), and laser ablation (*20*) amongst others.

**Data-guided discovery cycle**

Up until now selection of an elemental CSS composition is guided by chemical intuition. With the combinatorial explosion of possibilities that occurs when considering combinations and composition of constituent elements, it becomes almost impossible to discover optimal multinary electrocatalysts for specific reactions without an underlying theoretical framework. While no such framework exists for predicting the optimal set of constituent elements for a reaction, finding the most active composition of a given set is of equal importance. To accomplish this, we demonstrate a closed-loop data-driven high-throughput experimentation approach that iteratively combines (i) ab-initio simulations and modelling, (ii) combinatorial synthesis of materials libraries (MLs) and (iii) high-throughput characterisation, see Fig. 1. This starts with an initial hypothesis/prediction of a system of interest which is synthesized, in the form of thin-film MLs, comprising large compositional ranges which allow the efficient identification of property maxima and minima. The results are used to refine the model to improve predictions in the next cycle.

As a testbed, noble metals as electrocatalysts for the ORR allow robust and well-defined experimental parameters without risking material modification such as surface oxidation or dissolution. Moreover, established theoretical design principles exist for ORR performance, which have been confirmed experimentally. Theory and experiment suggest that an optimal catalyst binds *OH and *O with an adsorption energy 0.1 eV (*21*) and 0.2 eV (*22*) weaker than Pt, respectively. For comparing surfaces an estimate of the relative activity, *A*, is calculated from all binding energies with associated weights given by the probability of finding each type of site using Eq. 1:

$$A = \sum_{i=1}^{Z} \left( \prod_{k}^{metals} f_k^{n_{ik}} \right) \exp\left( -\frac{|\Delta E_i - \Delta E_{opt}|}{k_B T} \right) \qquad (1)$$

Per-site activity is assumed to increase exponentially with decreasing difference between the binding energy of an intermediate on site $i$, $\Delta E_i$, and the optimum given by the Sabatier principle, $\Delta E_{opt}$. $k_B$ and $T$ are the Boltzmann constant and temperature respectively. The per-site activity is then multiplied by the probability of site $i$ occurring, $\prod_k^{metals} f_k^{n_{ik}}$, where $k$ are constituent metals, $f$ are the fractions of $k$ in the overall composition, and $n_{ik}$ is the number of atoms of $k$ constituting the binding site. Finally, all per-site activities of the Z sites included in the model (Fig. S1) are summed to give the overall activity.

The computational challenge posed by CSS surfaces derives from the millions of possible local atomic arrangements, all having different binding energies defined by the composition and relative positions of atoms. A totally mixed alloy has a statistical distribution of different local atom arrangements which serve as active sites. Therefore, a fast method of calculating binding energies is needed. A DFT-dataset of thousands of binding energies enables fitting a model to describe the remaining millions of possible arrangements. This initial model predicted CSS activities of Ir-Pd-Pt-Rh-Ru for the ORR (*4*) and Pt-Pd-Cu-Ag-Au for the $CO_2$ and CO reduction reactions (*5*).

Here, we focus on the system Ag-Ir-Pd-Pt-Ru, with the constituents chosen based on the likelihood of forming a stable and active CSS, and a composition predicted with the initial model. For computational details of the DFT-calculated binding energies and regression method see SI text and Figs. S1 and S2. Linear regression was employed for all *OH and *O binding energy simulations. A most active composition was calculated using Eq. 1: $Ag_5Ir_5Pt_{20}Pd_{35}Ru_{35}$ assuming the *OH activity descriptor with 5-35% bounds on individual element compositions. These bounds were imposed to ensure a high probability for CSS stability.

Three Ag-Ir-Pd-Pt-Ru MLs (ML1, ML2 and ML3) covering different composition spaces centered around the predicted optimal compositions were fabricated using combinatorial co-sputtering from five elemental targets (Figs. S5 – S10) with composition ranges shown on Table S1. All 342 measurement areas (MAs, each 4.5 mm × 4.5 mm) on a ML are synthesized simultaneously in one experiment.

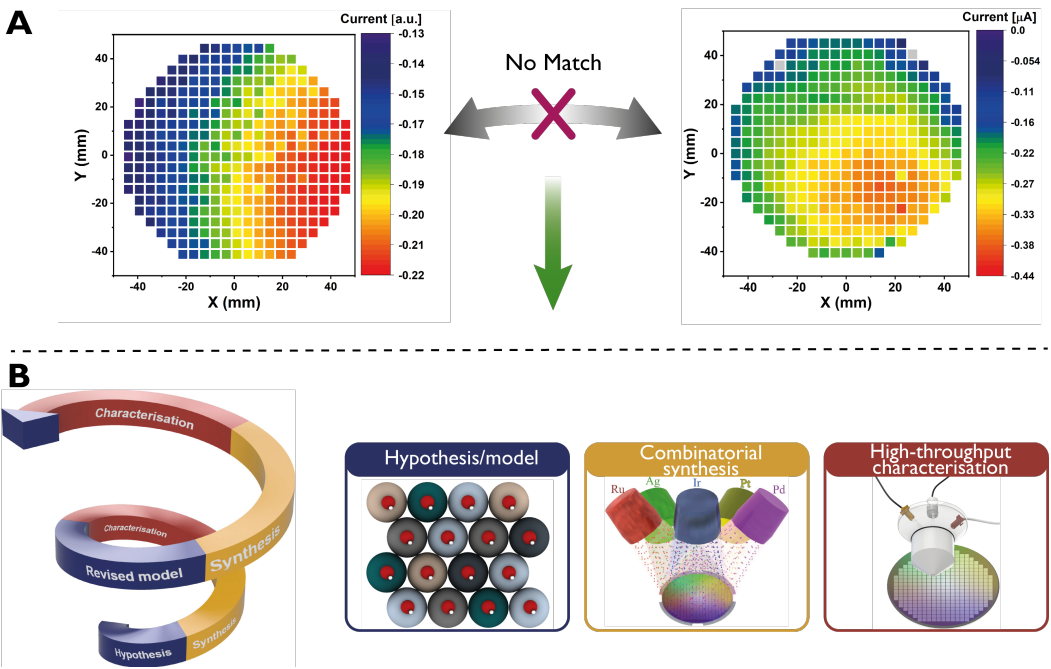

**Figure 1.** Schematic representation of the iterative materials discovery loop. **(A)** Predicted and experimentally obtained ORR activity on ML1 using SDC measurements where the initial model does not match the experimental result. **(B)** The data-driven discovery cycle combining prediction, combinatorial synthesis and high-throughput characterisation.

The MLs were investigated with scanning droplet cell (SDC) measurements in 0.1 M $HClO_4$. After rinsing with new electrolyte the SDC head was pressed onto each MA. The wetted area served as the working electrode. By measuring a linear sweep voltammogram (LSV) we evaluate the materials' ORR activity (see SI). Mapping composition-activity dependencies enables comparison with modelling. A homogeneous Pt thin film showed a small standard deviation with respect to the ORR activity for all MAs (Fig. S14), confirming reliability. Each ML was measured a second time, being turned by 90°, to rule out systematic errors (Fig. S15). Thus, fully consistent datasets unambiguously tied to the composition gradients (Fig. S16) were obtained.

**Simulating binding on a CSS surface**

Comparison between measurements and initial predictions shows that the activity trends are not predicted satisfactorily (Fig. 1A). Thus, we iteratively developed a series of models I, II, III built upon the initial model by additionally providing a measure of current per site, $j_i$, as a function of potential, $U$, see **Eq. 2**. Using the local composition around each binding site, (details seen in Fig. S1) surface binding energies are mapped. The Koutecký-Levich equation for rotating disk electrode (RDE) experiments is used to model $j_i$.

$$\frac{1}{j_i} = \frac{1}{j_{k_i}} + \frac{1}{j_D} \quad (2)$$

$$j_{k_i} = e^{\frac{-|\Delta E_i - \Delta E_{opt}| + \Delta E_{opt} - eU}{k_B T}} \quad (3)$$

Here $e$ is the charge on an electron. $j_i$ is calculated per site $i$, and then summed up over all sites, $N$, on the surface.

$$j_{tot} = \sum_i^N j_i = \sum_i^N \frac{1}{\frac{1}{j_{k_i}} + \frac{1}{j_D}} \qquad (4)$$

The magnitude of the per-site current response diminishes exponentially with increasing distance from the ORR volcano peak, similar to the initial model. The resulting shape of the current response can be correlated to the adsorption energy distribution pattern. The transport limitation is represented by $j_D$ and is set to 1. At low over-potentials there is little mass transport limitation in experiment, providing the basis for comparison with SDC measurements.

Three models were developed with distinct binding and site interactions as illustrated in Fig. 2. All models are 'brute force methods' since for every composition a small section of the surface is simulated (100 x 100 atoms, 3 layers) with Ag, Ir, Pd, Pt, and Ru randomly populating the face centred cubic (fcc) lattice. By extracting nearest neighbour compositions around on-top and hollow sites on this surface ~20,000 binding energies are predicted.

Model I focuses on the binding energy of *OH as a descriptor for catalytic activity, with the optimum energy being 0.1 eV weaker than on Pt. Assuming *OH binds to on-top sites without neighbouring site interactions (Fig. 2A left), using the binding energies on the surface (Fig. 2B) a measure of the current response at certain potentials (Fig. 2E) is calculated using Eq. 2. In contrast, model II considers *O binding to fcc-hollow sites rather than *OH on-top (Fig. 2A middle). In this case, even though neighbouring sites share an atom, no site interactions are assumed. The optimum energy is now 0.2 eV weaker than on Pt (*22*). The binding energies (Fig. 2C) are then used to predict current (Fig. 2F) as in model I. Applying models I and II to a single-element surface, scaling between fcc-hollow and on-top sites leads to an identical activity prediction since an adsorbate sees the same surface environment regardless of the binding site. This is not the case for a CSS since adsorption energy distributions for *OH and *O have different shapes, (Fig. 2B and C), confirming that fcc-hollow and on-top sites do not scale on a CSS. Interaction of *OH and *O at different binding sites is important due to the complex surface arrangements. Model III includes this effect and additionally considers that on a real CSS surface some sites will prefer an *O binding to the hollow position while other sites an *OH to on-top, and that sites can be blocked by neighbouring adsorption (Fig. 2A right). The simulated surface of 10,000 atoms is populated with *O and *OH adsorbates by filling the sites starting from the most stable on-top and hollow sites. Once the surface is filled completely only blocked sites remain vacant and are assumed to be inactive. Filled sites take part in the current calculation (Fig. 2G) as in models I and II. Since the surface is populated with neighbouring site blocking considered, the available binding energy distributions (Fig. 2D) are not simply a combination of the non-interacting *O and *OH distributions. Moving from a pristine surface to a situation where neighbouring sites interact, the binding energy distribution is altered.

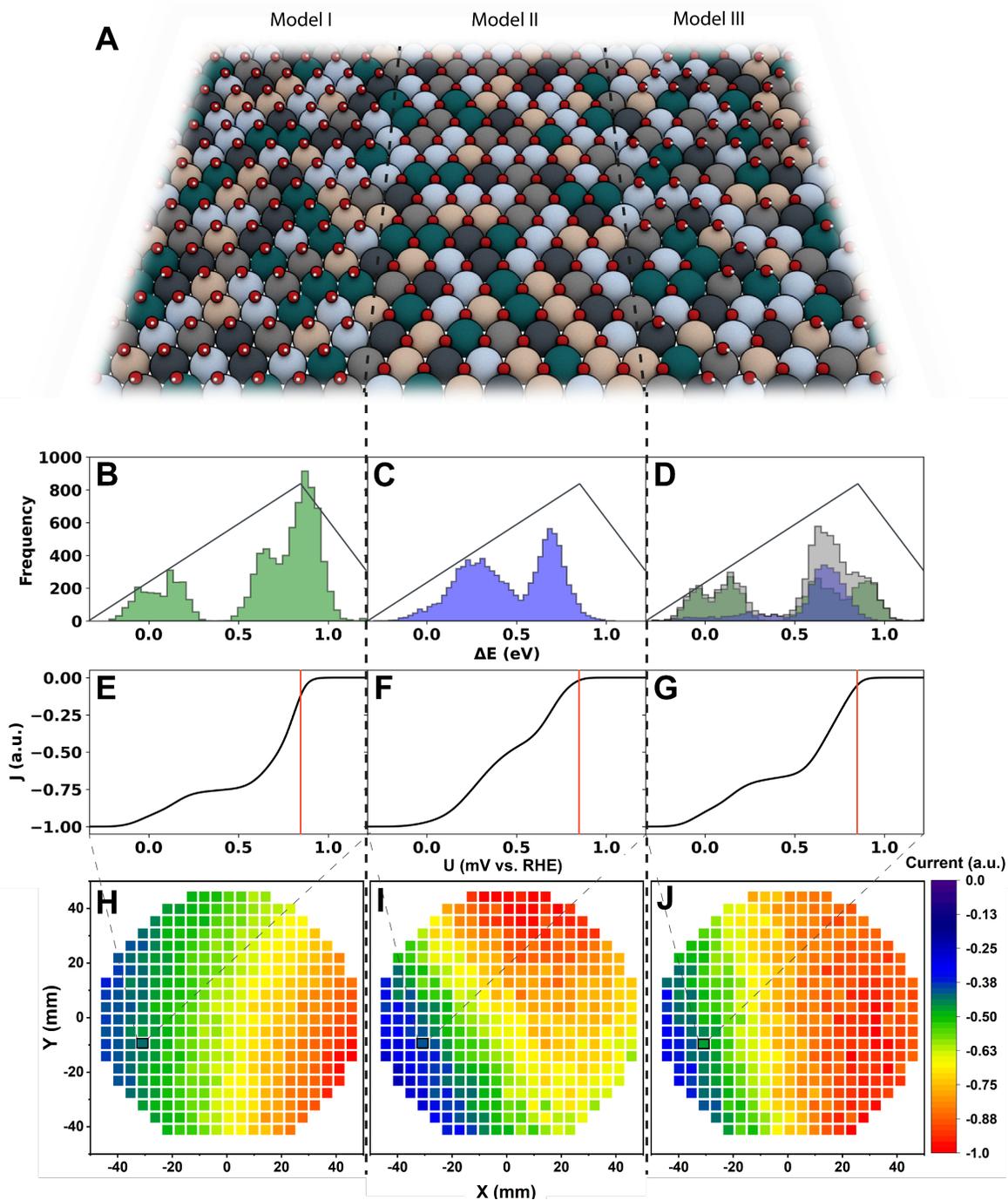

**Figure 2.** Comparisons between models I, II and III. **(A)** Schematic representation of how the CSS surface is populated in the models. Red atoms represent oxygen, white atoms represent hydrogen, and the other colours represent the CSS surface. **(B - D)** Histograms showing the binding energy distribution patterns of *OH (green), *O (blue), and combined (grey) on the simulated 10,000 atom surface. Volcano curves illustrate the optimum binding energy. **(E - G)** Example polarization curves for $Ag_4Ir_{16}Pd_{30}Pt_{14}Ru_{36}$ plotted as a function of potential. Red lines indicate the potential 820 mV vs. RHE. **(H - J)** Activity maps plotted using models I, II and III respectively. Current is calculated at 820 mV vs. RHE, compositions are taken from ML1. Selected compositions in **E** to **G** are indicated by the black box.

We demonstrate predictions by the three models, displaying predicted activity maps (Fig. 2 H-J) using composition data from ML1 (Energy-dispersive X-ray spectroscopy data shown in Figs. S5 to S10).

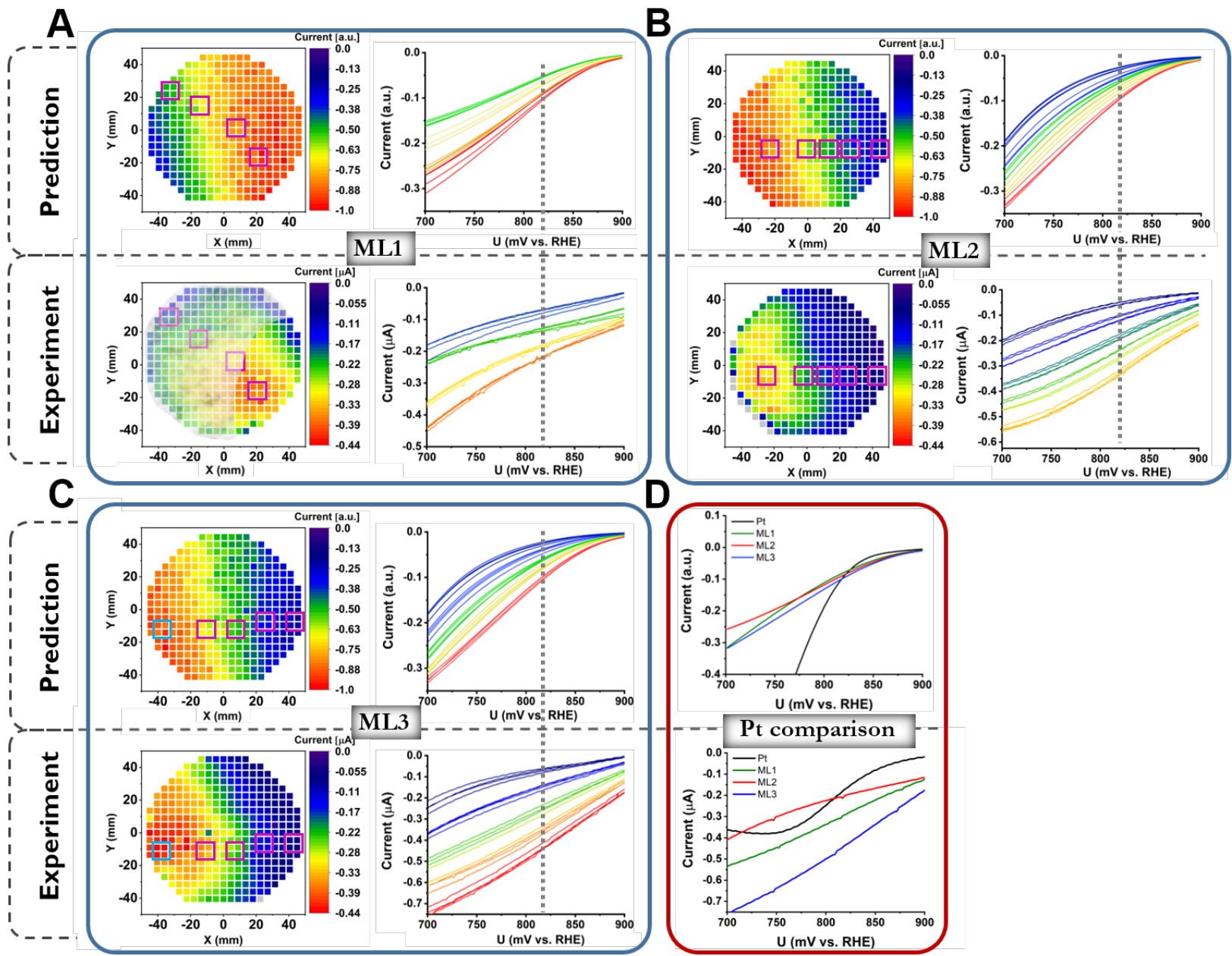

**Figure 3.** Comparison of predicted (model III) and experimental ORR activity trends for Ag-Ir-Pd-Pt-Ru MLs and Pt. (**A**) ML1, (**B**) ML2, and (**C**) ML3 activity maps with individual LSVs of the selected MAs covering high to low activity (indicated with the squares) are illustrated on the right. The current at 820 mV vs. RHE (indicated with the dashed line) was chosen as the measure of activity. ML1 shows larger differences when comparing predictions with experiment. This can be rationalised as all models exclusively consider an fcc structure. XRD measurements show that ML1 exhibits different phase regions: fcc, fcc + hcp, and hcp, (Fig. S11, greyed areas in ML1), whereas ML2 and ML3 are all fcc (Fig. S12 to S13). (**D**) LSVs of Pt thin film benchmark compared to the most active compositions of the MLs.

**ORR measurements**

Fig. 3 shows the comparison of predicted (model III) and experimental activity maps as well as measured LSVs for the ORR at selected MAs from regions with different activity (indicated with squares). All activity maps use the current (measured or calculated) at 820 mV vs. RHE as activity descriptor. Since the polarisation curves do not cross within the relevant potential range, the composition trends are not significantly affected by choosing different potentials. An already good match of predicted and measured data for ML1 is observed. XRD analysis revealed that ML1 consists of three regions with different crystal structure (see Fig. S11), which can interfere with the comparison and we highlighted the fcc-only region as the model assumes an fcc structure. For this reason, we then prepared additional MLs that were purely fcc (Fig. S12 to S13). They show a very good agreement with the activity trends. The simpler models I and II yield different composition trends (Fig. S3) emphasising the importance of the additional interaction used in model III.

In the comparison of the most active MA of each ML with the Pt polarisation curve we find that, especially in the relevant low overpotential region, Pt activity is surpassed by the best CSS compositions. Considering the drastically reduced Pt content in the CSS, it is shown that superior activities are achieved on CSS surfaces and the importance of implementing high-throughput iterative strategies to fully exploit their multidimensional search space is demonstrated.

**Conclusions**

To summarise, we combine simulation, machine learning, data-guided combinatorial synthesis and high-throughput characterisation to identify CSS compositions with high electrocatalytic activity, particularly for but not limited to the ORR. We demonstrate that models for predicting ORR activity based on simple descriptors do not contain enough information to make predictions on a CSS surface. This provides fundamental insights, namely that it is the interaction between adsorbates and resultant blocking that creates the active surface. We find a model incorporating the simplest adsorbate-adsorbate interactions is able to replicate experimental activity trends remarkably well. Comparison of data from over 1000 ORR activity measurements and machine learning guided predictions demonstrates an efficient methodology for high-throughput closed-loop materials design in the flourishing field of electrocatalysis on CSS surfaces.

**Acknowledgments:**

**Funding:** W.S. acknowledges funding from the Deutsche Forschungsgemeinschaft (DFG) under Germany´s Excellence Strategy (EXC 2033 – 390677874 – RESOLV) and from the European Research Council (ERC) under the European Union's Horizon 2020 research and innovation programme (grant agreement CasCat [833408]. A.L., A.S., and B.X. acknowledge funding from DFG projects LU1175/22-1 and LU1175/26-1, V.S. stipend from IMPRS SurMAT. T.B., J.P, C.C., and J.R. thank the Danish National Research Foundation Center fo High-Entropy Alloys Catalysis (CHEAC) DNRF-149, and from research grant 9455 from the VILLUM FONDEN.

**Author contributions:** T.B., B.X., T.L., and A.L. wrote the paper and all authors contributed to revisions. T.B., J.P., C.C., and J.R. developed the models. T.B. ran the DFT calculations. B.X. and A.S. synthesized and characterized the MLs. O.K., V.S., and T.L. performed electrochemical characterization. W.S., A.L., and J.R. guided the project.

**Competing interests:** Authors declare no competing interests.

# Supplementary Information

**Materials and Methods**

*Computational methods*

All calculations were set up in the Atomistic Simulation Environment (ASE) using GPAW as the DFT code with wave functions expanded as plane-waves with the chosen functional being RPBE. Adsorption energies used to train all models in the manuscript were calculated on 4 layered 2x2 atom slabs with periodic boundary conditions set in the x and y directions, a plane-wave energy cutoff of 400 eV and a Monkhurst-Pack k-point sampling of the Brillouin zone of (4,4,1). All adsorption energies used to verify model accuracy were calculated on 4 layered 3x4 atom slabs with the same periodic boundary conditions and energy cutoff but with Monkhurst-Pack k-point sampling of (3,3,1). Elements Ag, Ir, Pd, Pt and Ru were selected at random to occupy atom positions in each slab, with every slab being different. All slabs were relaxed to below 0.1eV/Angstrom per atom, using spin-paired orbitals, with 7.5 Angstrom of vacuum added above and below in the z-direction. The bottom 2 layers of all slabs were fixed geometrically to simulate a bulk structure, while the top 2 layers were allowed to relax. The lattice parameters of all slabs were calculated as the weighted average lattice parameter of the elements constituting the top layer. This was chosen to account for local strain effects on a CSS surface. The lattice parameters of pure elements were calculated for each pure element using a primitive fcc unit cell, and plane wave cutoff of 400 eV, Monkhurst-Pack k-point sampling of (8,8,8) and periodic boundary conditions in x,y and z directions. Energies of gas-phase reference molecules H2O and O2 were calculated and used to find adsorption energies using

$$\Delta E_{*OH} = E_{*OH} + \frac{1}{2}E_{H_2} - E_* - E_{H_2O} \quad (1)$$

$$\Delta E_{*O} = E_{*O} + E_{H_2} - E_* - E_{H_2O} \quad (2)$$

Where $\Delta E_{*OH}$ and $\Delta E_{*O}$ are the adsorption energies of *OH and *O, $E_{*OH}$ and $E_{*O}$ are the DFT energies of slabs with adsorbed OH and O respectively, $E_*$ is the DFT energy of the bare slab, $E_{H_2}$ and $E_{H_2O}$ are DFT energies of molecular references $H_2$ and $H_2O$ in gas phase, respectively.

Supervised machine learning was used to predict binding energies of *OH and *O on a simulated surface of 100x100 atoms, 3 atom layers thick. DFT calculations of 1341 *OH and 1976 *O binding energies on randomly populated 2x2 4 layered slabs served as the training set and DFT calculations of 40 *OH and *O binding energies on randomly populated 3x4 4 layered slabs served as the test set. The larger slabs were used to test the accuracy since, compared to the training set, they better represent the structural disorder found on a CSS surface. Linear regression (LR) was used, providing high accuracy adsorption energy predictions with test-set RMSD of 0.115 and 0.132 eV for *OH and *O respectively. The scikit-learn python package was used to perform the regression. The fingerprint used to encode the binding site element composition is shown on **Fig. S1**, where the binding site is split into different zones shown, with the prediction accuracy plots shown on **Fig. S2**.

*O and *OH DFT binding energies predicted on the simulated CSS surface were shifted to account for applied potential and the effect of interaction with water, zero-point and entropy corrections. *O and *OH binding energies have been normalized to Pt(111). With the *OH volcano peak being found to be 0.1 eV weaker than on Pt(111) [2012], the *O volcano peak is 0.2 eV weaker than *O on Pt(111) since responds twice as much to applied potential compared to *OH due to a double bond to the surface. To find an absolute scale for binding energies we utilize scaling between *OH and *OOH, where $E_{*OOH} = E_{*OH} + 3.2$ eV. By assuming that *OOH and *OH have the same interactions with water, zero-point and entropy corrections the relation likewise holds for adsorption free energies. With there being two charge transfer steps between *OH and *OOH the top of the *OH volcano lies at $U = (3.2 \text{ eV} - 2.46 \text{ eV})/2e = 0.86$ V where $e$ is the electron charge. Since the *OH volcano peak lies at $\Delta G_{*OH} = 0.86$ eV, the *O volcano peak lies at $\Delta G_{*O} = 2 \times 0.86$ eV due to the two-fold response to applied potential. When applying this to Pt(111) we find that $\Delta G_{*OH} = 0.76$ eV which is consistent with the result with interaction with water, zero-point and entropy corrections included [2004]. Since binding energies are normalized to Pt(111) we shift all predicted *OH binding energies by the difference between the calculated DFT binding energy and the binding free energy, $\Delta E_{*OH} - \Delta G_{*OH} = 1.017 \text{ eV} - 0.76 \text{ eV} = 0.257$ eV. *O DFT binding energies are halved to account for applied potential, leading to the volcano peaks for *OH and *O both lying at 0.86 eV.

Predicted activity maps for MLs 1, 2 and 3 are calculated using this model, as described in the main text. Comparisons between 3 models (I, II, and II) are shown on **Fig. S3.**

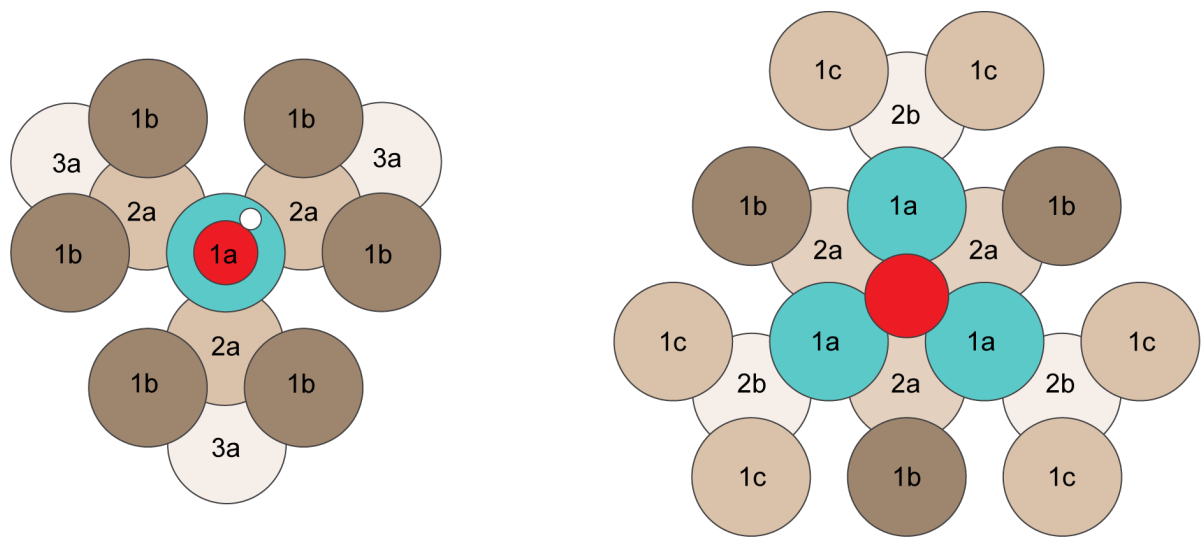

**Fig. S1.** Diagram of *OH on-top (left) and *O hollow (left) binding sites, with the nearest neighbour zones that are used to encode information for machine learning. The adsorbate atoms oxygen and hydrogen are red and white respectively. Atoms directly binding with the adsorbate are blue, and the other atoms have different colours to visually differentiate the zones. The number of atoms of each element in zones 1 (a, b, c), 2 (a, b) and 3 (a) for *OH on-top (and *O hollow) binding are converted into fingerprints and used as features. Here the layers are denoted by number (1$^{st}$ layer, 2$^{nd}$ layer and 3$^{rd}$ layer) and the distance from the binding site by letter (a closest, b next closest, c furthest). For five constituent elements this feature encoding allows the distinction of $Z = 5*210*35^2 = 1,286,250$ OH on-top sites and $Z = 35^4*210 = 315,131,250$ O hollow sites.

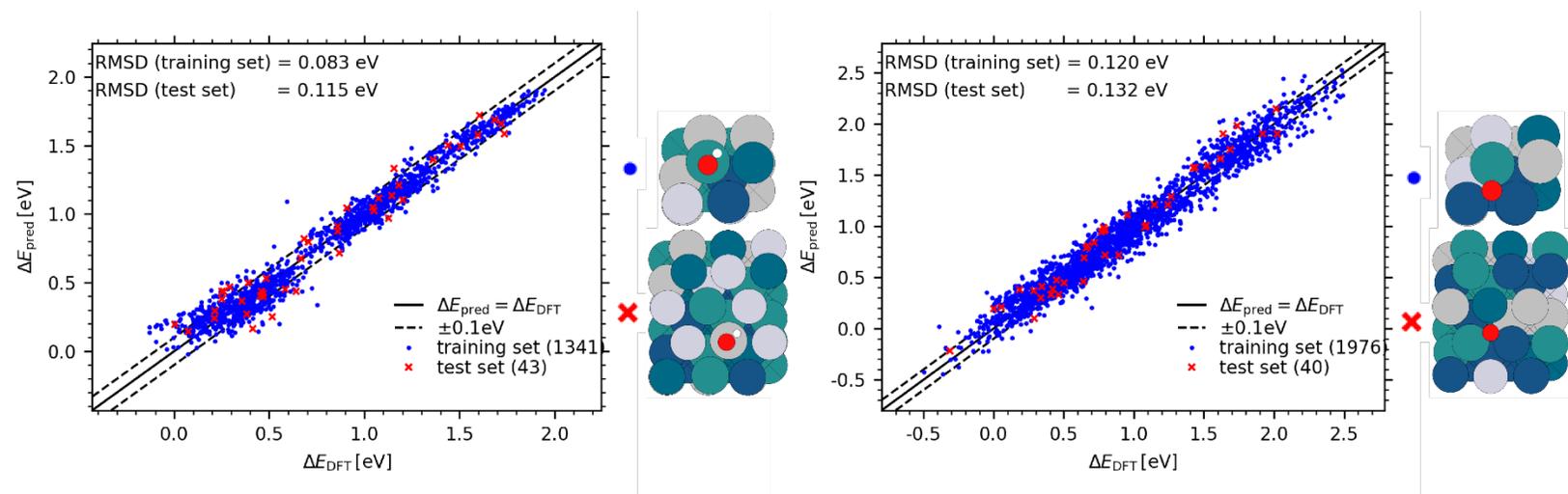

**Fig. S2.** Predicted vs. DFT scatter plots of binding energies of 1341 *OH on-top sites (left) and 1976 *O hollow sites (right). Linear regression was used for prediction, with root mean square deviation (RMSD) values for training (blue dots) and test (red dots) sets shown inset. Examples of the CSS slabs used for training and test are shown to the right of each graph.

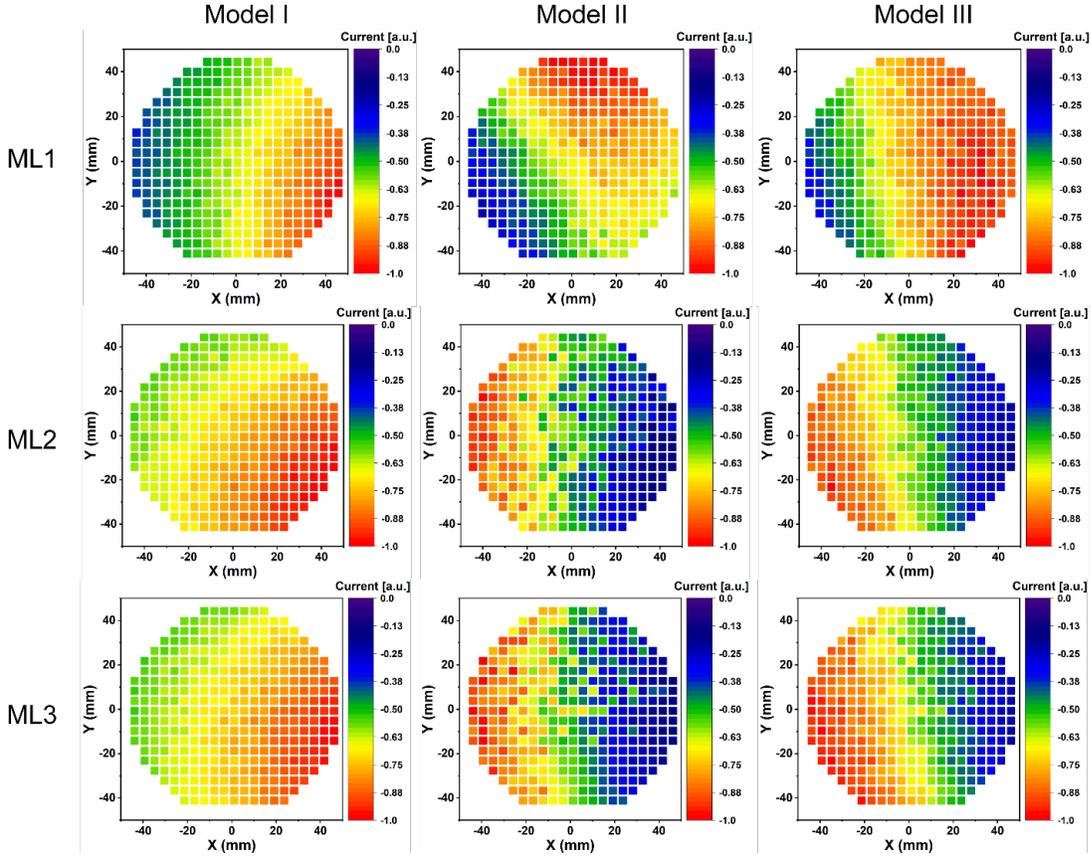

**Fig. S3.** Comparison of predicted activity maps on MLs 1, 2, and 3, using models I, II, and III, employing linear regression for binding energy predictions. Differences seen between models for each ML are due to the different assumptions, as visualised on **Fig. 2** of the main manuscript.

*Fabrication of thin-film MLs*

MLs comprise multinary thin-film alloys in an extensive and controllable multidimensional composition space. Magnetron sputter deposition of MLs was performed without intentional heating using five 100 mm diameter metallic targets: Ag (purity 99.99%), Ir (purity 99.99%), Pd (purity 99.99%), Pt (purity 99.99%) and Ru (purity 99.99%); 100 mm diameter sapphire wafers ($Al_2O_3$, Kyocera Corporation, $R_a$ = 20 nm) were used as a substrate for all depositions. To obtain a continuous compositional gradient, the substrate was kept stationary and each target was inclined with an angle of 45° with respect to the substrate. Prior to the deposition, the chamber vacuum was set on the order of $10^{-7}$ Pa. The MLs were deposited at a pressure of 2.7 Pa with Ar flow of 80 sccm. **Table S1** lists all process parameters. The predicted composition was deposited in the center of the ML by adjusting the sputter rates of the five targets. The gradients of each element over a ML is shown in **Fig. S15**.

**Table S1.** Sputter parameters for the ML1, ML2 and ML3, respectively. (DC: direct current, RF: radio frequency)

| material library | deposition power (W) | | | | | deposition duration (s) | composition range |
| --- | --- | --- | --- | --- | --- | --- | --- |
| | Ag [RF] | Ir [DC] | Pd [RF] | Pt [RF] | Ru [DC] | | |
| ML1 | 57 | 32 | 149 | 192 | 133 | 390 | $Ag_{1-9}Ir_{8-18}Pd_{17-49}Pt_{12-33}Ru_{17-52}$ |
| ML2 | 13 | 28 | 252 | 185 | 42 | 360 | $Ag_{0-3}Ir_{6-20}Pd_{44-73}Pt_{13-32}Ru_{4-20}$ |
| ML3 | 16 | 28 | 252 | 237 | 42 | 360 | $Ag_{0-3}Ir_{4-18}Pd_{42-71}Pt_{14-41}Ru_{4-19}$ |

*High-throughput characterisation of MLs*

*Roughness of CSS MLs and Pt*

Topographical images of the ML1, ML2 and ML3 and a Pt thin film were measured by atomic force microscopy (AFM, Bruker Dimension Fastscan) using Fastscan mode. For surfaces, whose roughness is characterized by a single length scale, roughness parameters were calculated by the arithmetic mean roughness $R$a. ML1, ML2 and ML3 have comparable surface roughness with that of the Pt benchmark thin film, as shown on **Fig. S4.**

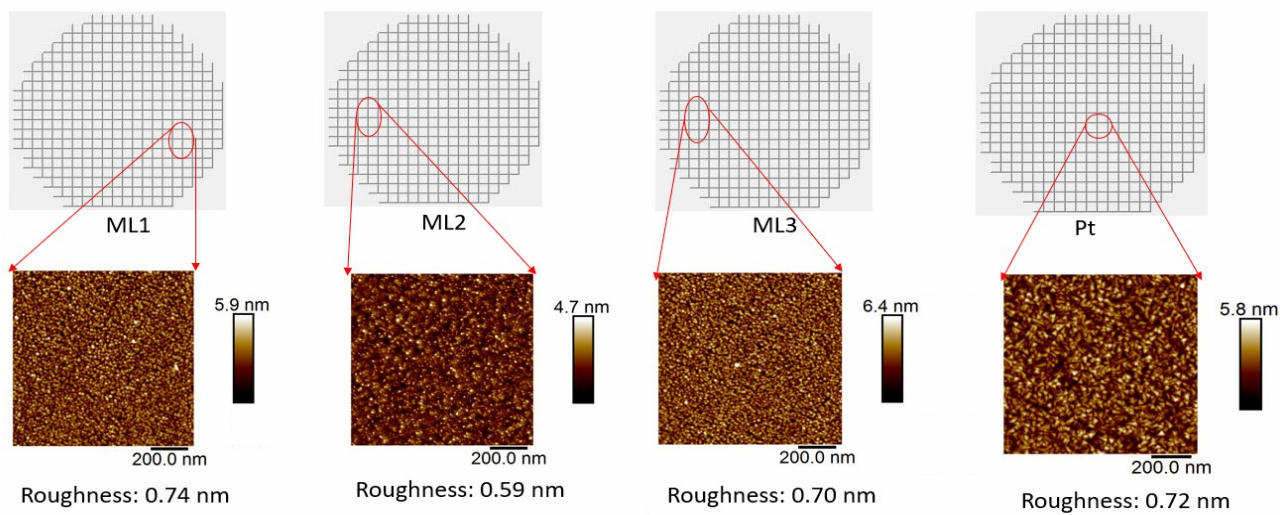

Roughness (mean of the absolute height) $R_a = \frac{1}{L}\int_0^L |y|\,dx$

**Fig. S4.** AFM images and extracted roughness values for the MLs and Pt.

*Compositional characterisation of MLs*

The elemental compositions of all MAs in the MLs were determined using automated energy dispersive X-ray spectroscopy (EDX) at 20 kV acceleration voltage in a scanning electron microscope (SEM, JEOL 5800) using a detector (INCA X-act, Oxford Instruments). **Figs. S5, S7** and **S9** show color-coded composition gradients for ML1, ML2, and ML3 respectively. These composition data are shown in **Figs. S6, S8** and **S10** in the form of color-coded pie charts.

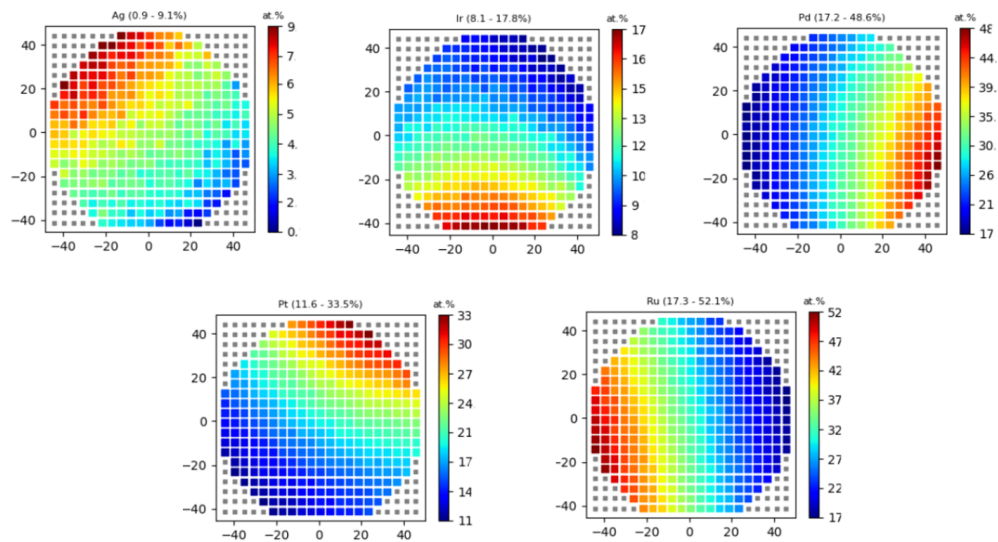

**Figure S5.** Composition ranges for ML1: Ag(1 – 9%), Ir(8 – 18%), Pd(17 – 49%), Pt(12 – 34%), Ru (17 – 52%).

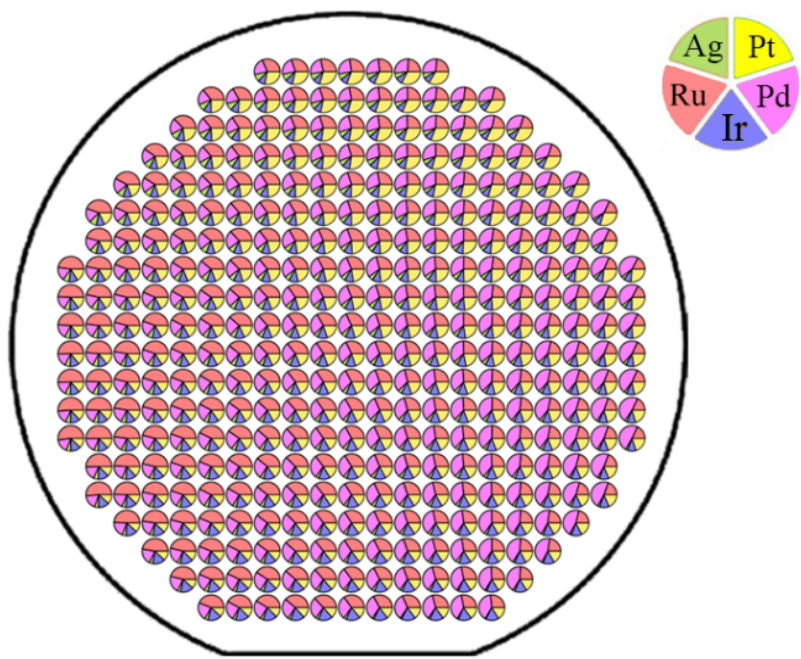

**Figure S6.** Pie-chart diagram for ML1, which indicates the relative elemental compositions at each of the 342 MAs. Legend shows the locations of elements.

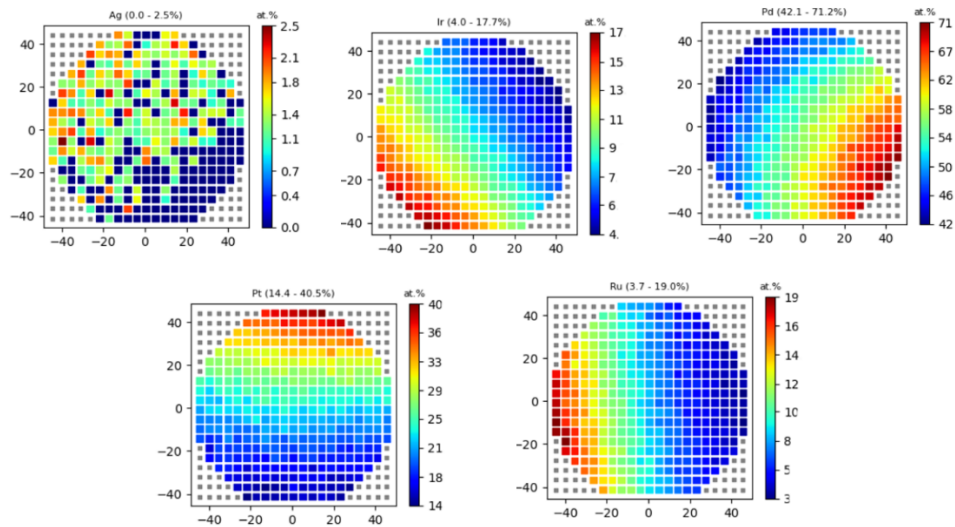

**Figure S7.** Composition gradients for ML2: Ag(0 – 3%), Ir(6 – 20%), Pd(44 – 73%),Pt(13 – 32%), Ru (4 – 20%).

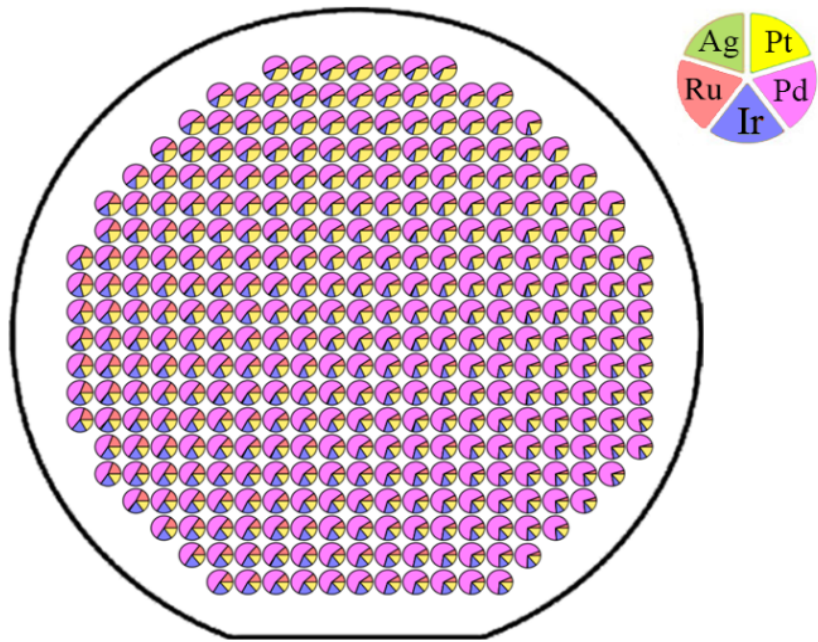

**Figure S8.** Pie-chart diagram for ML2, which indicates the relative elemental compositions at each of the 342 MAs. Legend shows the locations of elements.

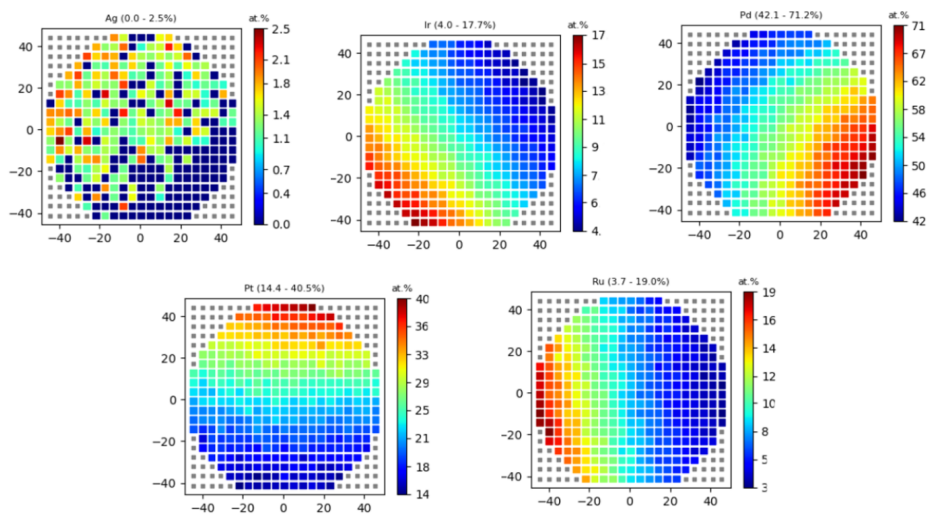

**Figure S9.** Composition ranges for ML3: Ag(0 – 3%), Ir(4 – 18%), Pd(42 – 71%), Pt(14 – 41%), Ru (4 – 19%).

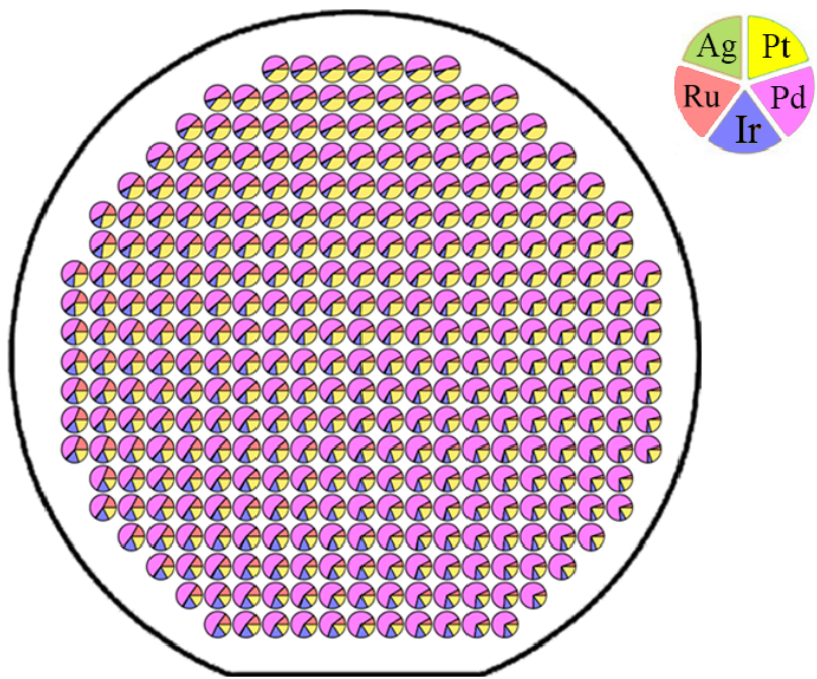

**Figure S10.** Pie-chart diagram for ML3, which indicates the relative elemental compositions at each of the 342 MAs. Legend shows the locations of elements.

*Crystallographic characterization of MLs*

High-throughput crystallographic phase analysis was performed using X-ray diffraction (XRD) in Bragg−Brentano geometry (Bruker D8 Discover, equipped with a VANTEC-500 area detector, Cu Kα radiation, sample to detector distance = 149 mm; collimator diameter = 1 mm). The acquired XRD patterns were integrated into one-dimensional data sets (diffracted intensity in dependence of the diffraction angle 2θ) and were then used for phase identification. The phases were identified by comparing the measured patterns with references from the Inorganic Crystal Structure Database (ICSD). **Figs. S11** to **S13** show these data for ML1 to ML3.

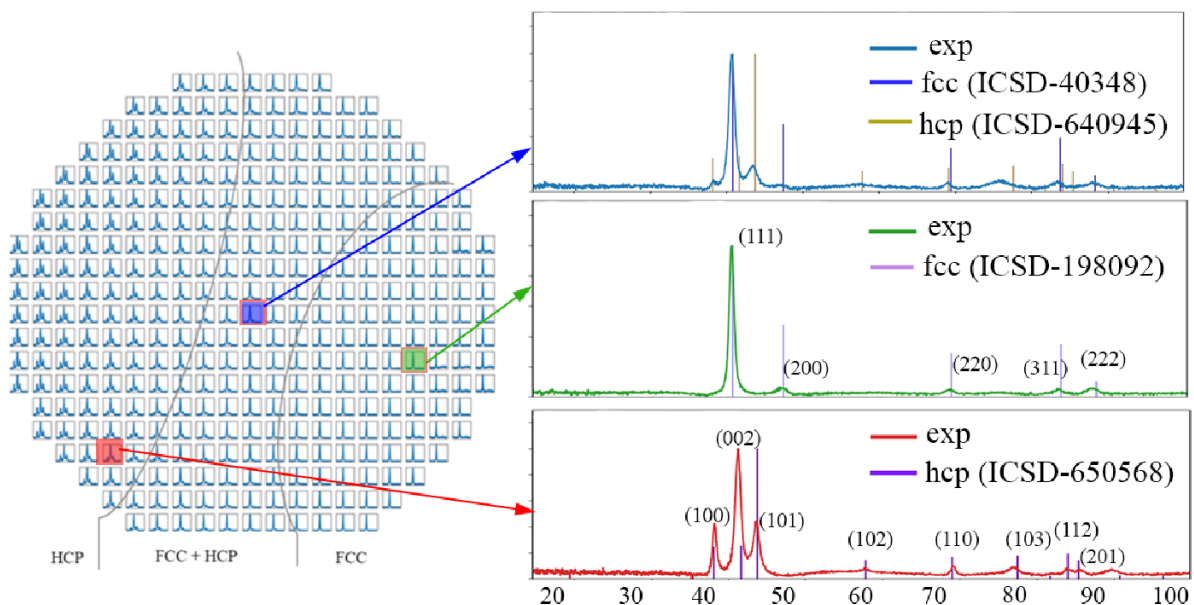

**Figure S11.** Phase and XRD map of the as-deposited Ag-Ir-Pd-Pt-Ru ML1. ML1 comprises three phase regions: pure hcp (left), fcc + hcp (middle) and pure fcc (right).

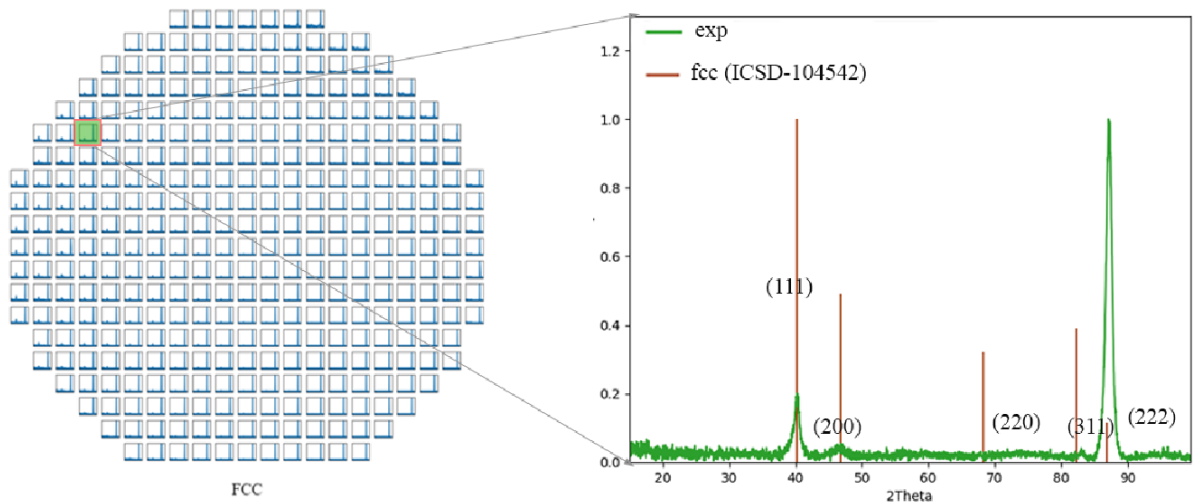

**Figure S12.** Phase and XRD map of the as-deposited Ag-Ir-Pd-Pt-Ru ML2. At the site of each MA the measured XRD pattern is plotted. All films of ML2 are crystalline and all 342 MAs show pure fcc phase patterns, as matched to ICSD-104542.

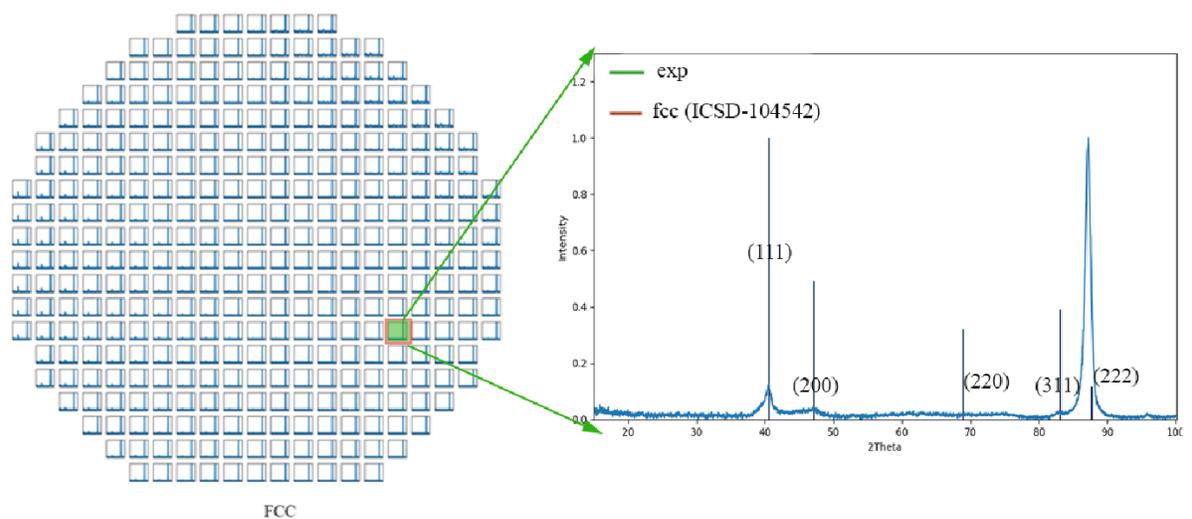

**Figure S13.** Phase and XRD map of the as-deposited Ag-Ir-Pd-Pt-Ru ML3. ML3 is crystalline and all 342 MAs are pure fcc.

*Electrochemical characterisation*

All MAs of the MLs were analysed using a high-throughput scanning droplet cell (SDC), which allows localised characterisation over the ML. All electrochemical measurements were conducted in 0.1 M HClO$_4$ electrolyte in a three-electrode system containing a Ag|AgCl|3M KCl and Pt wire as a reference and counter electrode, respectively. Linear sweep voltammetry was performed with a scan rate of 10 mV s$^{-1}$. All potentials are reported versus the RHE according to the following equation:

$$U_{RHE}(V) = U_{Ag|AgCl|3M\ KCl} + 0.210 + (0.059 \times \text{pH}) \quad (3)$$

where U$_{(Ag|AgCl|3\ M\ KCl)}$ is the potential measured vs. Ag|AgCl|3M KCl reference electrode, 0.210 V is the standard potential of the Ag|AgCl|3M KCl reference electrode at 25 °C. 0.059 refers to a temperature of 298 K.

The small electrolyte reservoir and the tube-type connection inside the SDC head allow the accurate determination of the voltammetric response in the kinetic region of the ORR at small overpotentials. The linear sweep voltammetry ORR response at each of the 342 MAs on a ML was automatically measured allowing the extraction of catalytic activity.

*Confirmation of reliable experimental activity trends*

The automated setup preserves identical measurement conditions, thus the electrocatalytic ORR activity can be compared between the MAs of a ML. To rule out any systematic errors in activity trends induced by the SDC setup, a pure Pt thin film was measured under the same conditions, which serves as benchmark catalyst for the ORR and confirms the close overlap of the individual LSV curves (**Fig. S14**). Since the variation between the different curves is insignificant compared to the current differences observed for the different composition MAs of the CSS thin film libraries shown in the main manuscript, the observed trends can be assigned to the composition effect.

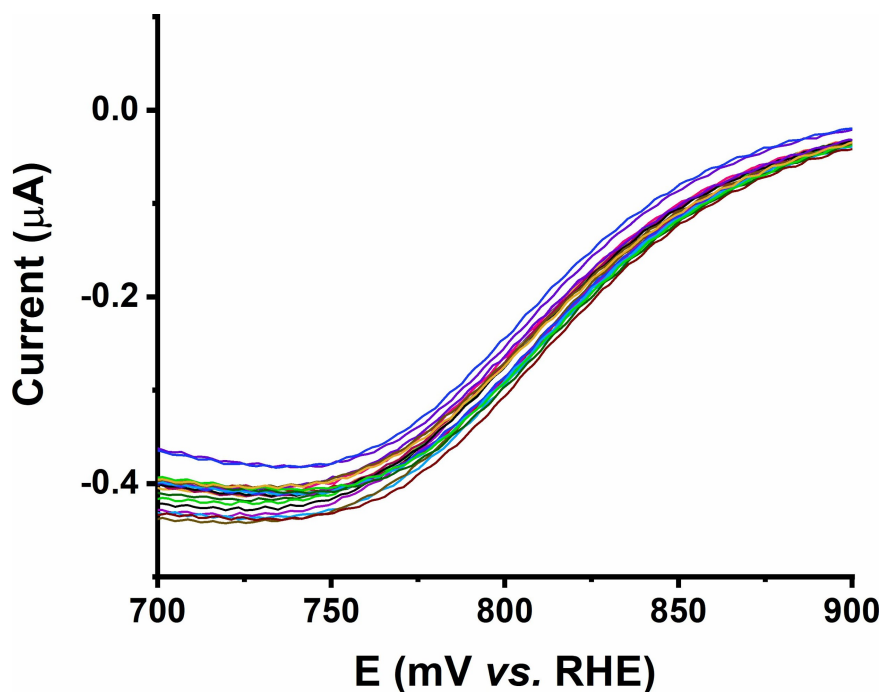

**Figure S14.** SDC measurement of a pure Pt thin film (deposited on 100 mm diameter substrate identical to those used for MLs). The exemplary polarisation curves of multiple MAs confirm close overlap, whose difference is insignificant compared to the variation of the Ag-Ir-Pt-Pd-Ru CSS thin-film MLs. Different colours are arbitrary and chosen to assist distinguishing the individual curves.

Additionally, each ML was measured a second time with 90° rotation. The induced different measurement order for the MAs guarantees that systematic trends in x or y direction or trends over time can be observed. The obtained activity trends are also shifted by 90° (**Fig. S15**), confirming that these must be caused by the composition gradient as the only constant variation for each MA and the trends can be unambiguously related to the composition effect.

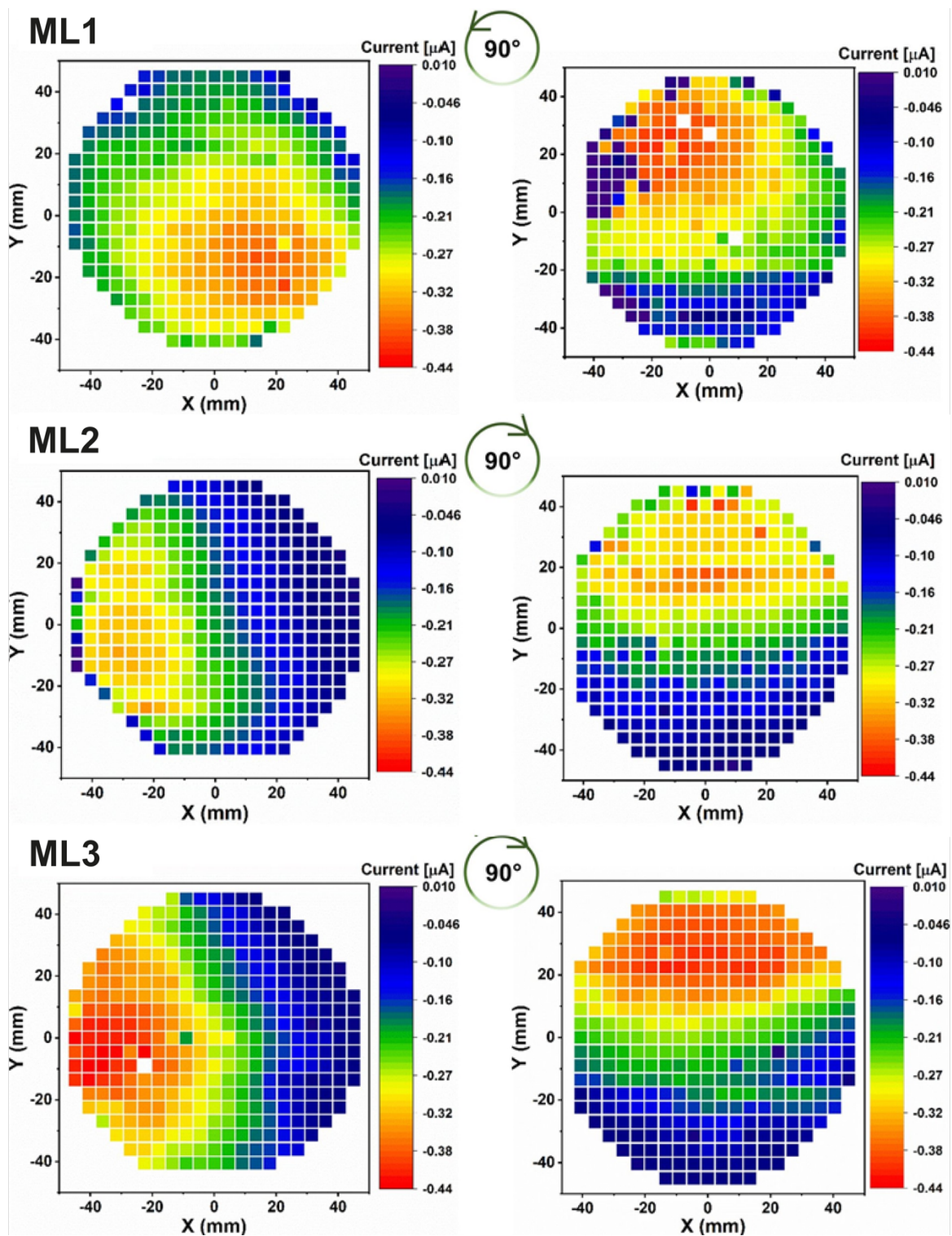

**Figure S15.** Confirmation of reliable activity trends, which are only caused by compositional gradients by repetition of SDC measurements for each ML in a 90° rotation.

*Correlation of activity trend and elemental content trend*

To understand the complex correlation between composition and activity, we analysed the correlation between composition and activity across one line of the MLs spanning high to low activity (**Fig. S16**). Following the line as represented by the x coordinate, a linear trend in element content for each element is obtained as expected based on the linearity of distance to the element target during the sputter process. Analogous plotting of the current at 820 mV vs. RHE as the activity descriptor suggests a linear correlation as well with reaching an optimum in between two reversed linear trends.

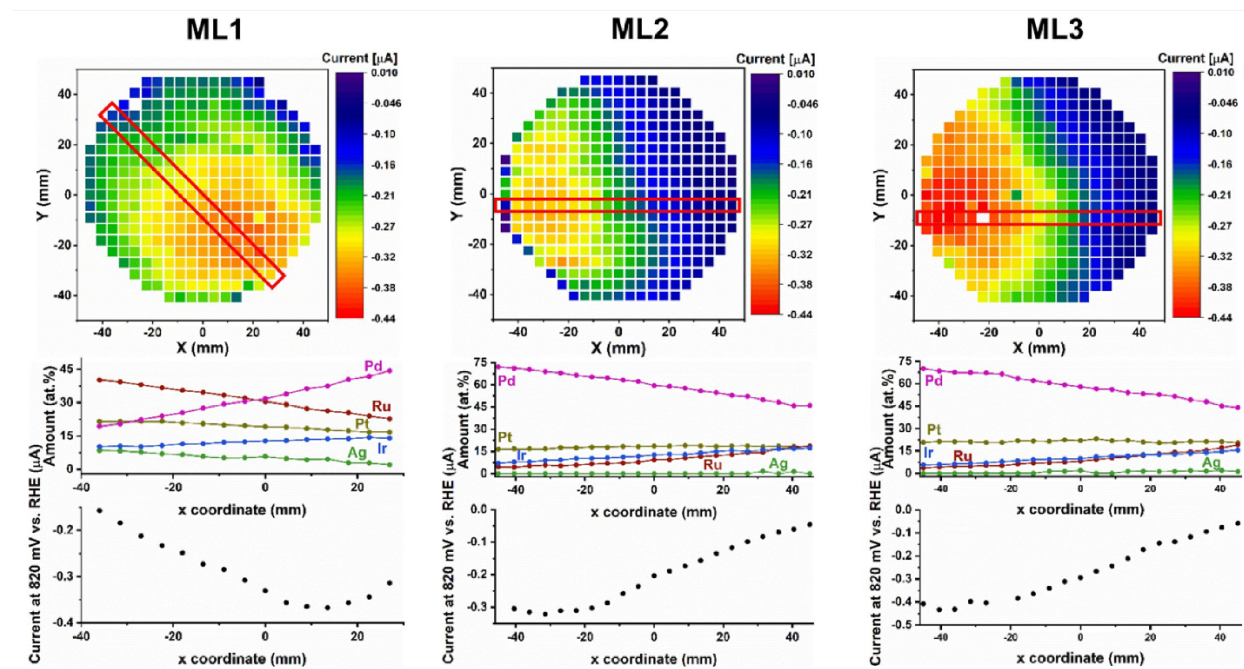

**Figure S16.** Correlation of molar ratio of each element (middle row) at each of the MAs highlighted with the blue arrow of the MLs (top row) and the related activity represented by the current at 820 mV vs. RHE (bottom row). The lines in the activity maps indicate the MAs used for this analysis, defined by their x coordinate.